\documentclass[twocolumn,prb,showpacs]{revtex4}
\usepackage{amsmath}
\usepackage{amssymb}
\usepackage{graphicx}
\usepackage{bm}

\begin{document}

\title{Large displacement strain theory and its application to graphene}

\author{J. A. Crosse}
\email{alexcrosse@gmail.com}
\affiliation{Department of Electrical and Computer Engineering, National University of Singapore, 4 Engineering Drive 3, Singapore 117583.}

\date{\today}

\begin{abstract}
Under the application of a force, a material will deform and, hence, the crystal lattice will experience strain. This induced strain will alter the electronic properties of the material. In particular, strain in graphene generates an  artificial vector potential which, if spatially varying, admits a pseudo-magnetic field. Current theories for spatially varying strain use linear or finite strain theory whose derivation is based on small displacements of infinitesimal length vectors. Here we apply a differential geometry method to derive a strain theory for large displacements of finite length vectors. This method gives a finite displacement term whose contribution is comparable to that of the linear strain term. Further to this, we show that a `domain wall'-like pseudo-magnetic field profile can be generated when a wide graphene ribbon is subjected to a pair of opposing point forces (point stretch). The resulting field is a function of the new finite displacement term only and displays a maximum strength of over three times that which is predicted by the linear strain theory. These results extend the current theories of strain, which are based on the transformation of infinitesimal length vectors, to finite length vectors, thus providing an accurate description of pseudo-magnetic field structures in strained materials.
\end{abstract}


\maketitle
%%%%%%%%%%%%%%%%%%%%%%%%%%%%%%%%%%%%%%%%%%%%%%%%%%%%%%%%%%%%%%%%%%%%%%

\section{Introduction}

Extensive studies of graphene have shown that it displays a wide range of remarkable electronic  properties \cite{rev1, rev2, rev3} and there has been much speculation on its role in future electro- and electro-optical devices. However, although unique, the properties of graphene are not ideal. For example, the gapless band structure and resulting minimum conductivity leads to low on-off ratios, a major obstacle in the development of usable graphene transistors \cite{trans}. Thus, in order to create efficient devices, one needs to alter graphene's intrinsic electronic properties. A number of approaches to this `band structure engineering' have been investigated; geometric confinement \cite{confine1, confine2}, doping \cite{dope1, dope2}, and substrate interaction effects \cite{substrate1, substrate2}, to name but a few. 

Another intriguing possibility is the use of strain. The band structure of a material is directly related to its crystal lattice. By applying a force, one can deform a material's crystal lattice and, hence, change its electronic properties. A number of studies have already proposed, gap generation \cite{pereiraTBM}, modification of graphene's optical properties \cite{pereiraopt1, pereiraopt2} and even the appearance of superconductivity \cite{supercon} under strain. Another feature of strained graphene is the appearance of artificial vector potentials and pseudo-magnetic fields \cite{pereiraA,pereiraPRL,geimhall,peeters1,peeters2,peeters3,peeters4,error} with predicted field strengths ranging from tens \cite{geimhall} to thousands of Tesla \cite{peeters1, peeters2,peeters3,peeters4} for both in-plane and out-of-plane strains. The ability to generate such pseudo-magnetic fields would remove the need for cumbersome external field generation and could pave the way for a generation of highly compact magneto-electronic and spintronic devices. In light of this, there have been a number of previous studies using a variety of methods; rotation free linear strain theory \cite{pereiraTBM,pereiraA}, rotational finite strain theory \cite{error} or, more recently, rotational finite strain theory up to second order in the strain tensor \cite{peeters2} as well as first principle atomistic simulations \cite{peeters1,peeters3}. However, both the infinitesimal (linear) and finite strain tensors are based on the small displacements of infinitesimal length vectors \cite{LandL7}. It would be advantageous to generalize this to large displacements of finite length vectors.

Here, we present a theory that uses a differential geometry method from which one is able calculate the exact, strain induced displacements for finite length vectors, thus going beyond the finite strain theory, which is based on small displacements of infinitesimal length vectors. This method gives a new finite displacement term that is found to be of comparable magnitude to the rotation and linear strain contributions. As an example, we compute the pseudo-magnetic field generated in graphene by a `point stretch' - where a pair of opposing point forces act laterally across the centre of a wide graphene ribbon. The resulting field is found to be a function of the finite displacement term only and that it displays a `domain wall'-like profile with a maximum field strength that is over three times that which is predicted by the linear strain theory alone.

\section{General Theory of Deformations}

In a strained material, the locations of the constituent atoms change from their equilibrium position, $R_{i}$, to $R'_{i} = R_{i} + u_{i}(\textbf{R})$, where $u_{i}(\textbf{R})$ is the displacement vector, which is, itself, a function of the equilibrium position of the atom (in the following we will use index notation with implied summation over repeated indices). In general, for a given applied force, the displacements, $u_{i}(\textbf{R})$, are not (\textit{a priori}) known and, hence, one characterizes the deformation in terms of the strain tensor, $\varepsilon_{ij}$, which can be found from the stress tensor, $\sigma_{ij}$, using the generalized Hooke's Law, $\varepsilon_{ij} = S_{ijkl}\sigma_{kl}$. (In principle the displacements can be found from the elastic Green function, however, formulating the equations of equilibrium for a material body usually involves some assumptions about the form of the strain tensor \cite{LandL7}). The rank-4 tensor $S_{ijkl}$ is the compliance tensor whose components are related to the mechanical properties of the material. Previous studies \cite{pereiraTBM, pereiraA}, which have considered rotation free, linear displacements (symmetric, spatially constant strain tensors), have found that the locations of the displaced atoms are given by, $R'_{i} = (\delta_{ij} + \varepsilon_{ij})R_{j}$, where $\delta_{ij}$ is the Kronecker delta. Rotations are easily accounted for by the inclusion of the rotation tensor \cite{peeters2, error}, $\omega_{ij}$, via $R'_{i} = (\delta_{ij} + \varepsilon_{ij} + \omega_{ij})R_{j}$. The rotation tensor is antisymmetric, trace free and related to the strain tensor by $\bm{\nabla}\times\bm{\omega} = -\bm{\nabla}\times\bm{\varepsilon}$.

\begin{figure}[h]
\centering
\includegraphics[width=0.9\linewidth]{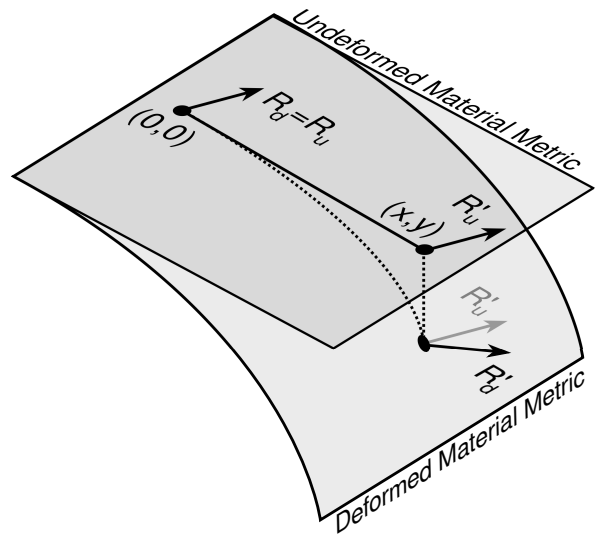}
\caption{Schematic diagram of parallel transport. The vector $\mathbf{R}_{u}$ in the undeformed material is unchanged when parallel transported along the solid line to $(x,y)$. The vector $\mathbf{R}_{d}$ in the deformed material is changed when parallel transported along the dotted line to $(x,y)$. This change is a direct result of the spatially varying displacement of the material and, hence, from this change the displacements can be found.}
\label{metric}
\end{figure}

The inclusion of large, spatially varying displacements is slightly more involved. The infinitesimal strain tensor gives the local change of an infinitesimal length over small displacements. For the change of finite lengths over a large displacements one needs to `integrate' the infinitesimal strain tensor over the deformation. This can be done using methods from differential geometry to account for the change in the strain tensor as the deformation progresses. The change in the element of length of an infinitesimal vector, $dR_{i}$, under strain is given by \cite{LandL7}
\begin{equation}
dl^{2} = g_{ij}dR_{i}dR_{j} = (\delta_{ij}+2\varepsilon_{ij})dR_{i}dR_{j}.
\label{gij}
\end{equation}
Thus, the strain tensor acts as a metric with the large, spatially varying displacements accounted for by the `non-euclidean' nature of the strain tensor. Note that, since the metric is, by definition, an infinitesimal object, one only needs consider the unique infinitesimal strain tensor and not one of the many finite strain tensor (e.g. Green-Lagrange, Almansi, etc.). For vanishing strain, the metric is euclidean, $g_{ij} = \delta_{ij}$, and hence the tangent spaces at different points on the manifold are identical. Thus, a vector, $\mathbf{R}_{u}$, defined in the tangent space at the origin and a vector, $\mathbf{R}'_{u}$, defined in the tangent space at a point $(x,y)$ are comparable and the parallel transport of $\mathbf{R}_{u}$ from the origin to $(x,y)$ leaves it unchanged ($\mathbf{R}_{u}$ = $\mathbf{R}'_{u}$). For non-vanishing strain the `non-euclidean' nature of the metric means tangent spaces at each point are different. Thus, a vector, $\mathbf{R}_{d}$, defined in the tangent space at the origin and a vector, $\mathbf{R}'_{d}$, in the tangent space at a point $(x,y)$ are not comparable. The parallel transport of $\mathbf{R}_{d}$ from the origin to the $(x,y)$ causes it to change by an amount proportional to the metric connections. Thus, in general, $\mathbf{R}_{d} \neq \mathbf{R}'_{d}$ with the difference between the two dependent on the form of the metric (which in this case is a function of the strain tensor). As the tangent space at the origin in the deformed material is isomorphic to euclidean space and, hence, the undeformed material, we have $\mathbf{R}_{u} = \mathbf{R}_{d}$. Thus, the parallel transport of $\mathbf{R}_{u}$ from the origin to the point $(x,y)=(R_{u,x},R_{u,y})$ will give $\mathbf{R}'_{d}$ and hence the displacement owing to the spatial variation of the strain tensor (See Fig. \ref{metric}).
To compute this we employ the parallel transport equation, familiar from differential geometry,
\begin{equation}
\frac{DR'_{i}}{D\lambda} = -\Gamma_{ijk}[\mathbf{R}(\lambda)]R_{j}(\lambda)\frac{dR_{k}(\lambda)}{d\lambda},
\end{equation}
where $\lambda$ parametrizes the path over which the vector is parallel transported. Here, $\Gamma_{ijk}$ are the metric connections that can be expressed uniquely in terms of the Christoffel symbols of the first kind and are give, in terms of the metric, by
\begin{equation}
\Gamma_{ijk} = \frac{1}{2}\left(\frac{\partial g_{ij}}{\partial x_{k}}+\frac{\partial g_{ik}}{\partial x_{j}}-\frac{\partial g_{jk}}{\partial x_{i}}\right).
\label{con}
\end{equation}
We wish to transport the vector $R_{i}$ from the origin at $(0,0,0)$ along it's length to $(R_{x},R_{y},R_{z})$. Thus, we use the parametrization $R_{i}(\lambda) = \lambda R_{i} = (\lambda R_{x},\lambda R_{y},\lambda R_{z})$ where $\lambda \in [0,1]$ and $dR_{i}/d\lambda = R_{i}$. Thus the spatially varying displacement is given by
\begin{equation}
R'_{i} = -R_{j}R_{k}\int_{0}^{1} d\lambda\,\lambda\Gamma_{ijk}[\mathbf{R}(\lambda)].
\end{equation}
Essentially, we have integrated the infinitesimal strain tensor over the deformation and hence have found the displacement of finite length vectors. This goes beyond the usual finite strain theory which, although second order, is still based on the displacement of infinitesimal vectors. Note that, in the case of linear deformations (constant strain tensor), the Christoffel symbols, $\Gamma_{ijk}$, vanish. Thus, the finite displacement term also vanishes and one recovers the results of previous studies \cite{pereiraTBM, pereiraA}.

Finally, one finds that the change in the vector locations of atoms displaced by a general spatially varying strain are given by
\begin{equation}
R'_{i} = \left(\delta_{ik}+\varepsilon_{ik}+\omega_{ik}\right)R_{k} - \Sigma_{ijk}R_{j}R_{k}.
\label{def}
\end{equation}
where $\Sigma_{ijk} = \int_{0}^{1} d\lambda\,\lambda\Gamma_{ijk}[\mathbf{R}(\lambda)]$ is the finite displacement term. Hence from Eq. \eqref{def} one is able to find the displacement vectors, $u_{i}(\mathbf{R})$, for any point in the deformed material. One should note that this expression is only valid for elastic deformation.

\section{Bond Deformation}

The electronic properties of a material are determined (to first approximation) by the relative locations of neighbouring atoms. Under strain the atoms are displaced and hence these interatomic distances are changed. Using the expression for a general displacement in Eq. \eqref{def}, the change in the relative distance between two atoms at $\mathbf{R}_{n}$ and $\mathbf{R}_{m}$ is given by
\begin{align}
R'_{n,i} - R'_{m,i} &= [R_{n,i}+\varepsilon_{ik}(\mathbf{R}_{n})R_{n,k}+\omega_{ik}(\mathbf{R}_{n})R_{n,k} \nonumber\\
&\hspace{3cm} - \Sigma_{ijk}(\mathbf{R}_{n})R_{n,j}R_{n,k}]\nonumber\\
&- [R_{m,i}+\varepsilon_{ik}(\mathbf{R}_{m})R_{m,k}+\omega_{ik}(\mathbf{R}_{n})R_{n,k}\nonumber\\
&\hspace{3cm} - \Sigma_{ijk}(\mathbf{R}_{m})R_{m,j}R_{m,k}].
\end{align}
Defining the interatomic distance as $\mathbf{R}_{\alpha} = \mathbf{R}_{n} - \mathbf{R}_{m}$ and employing the symmetries of the Christoffel symbols, $\Gamma_{ijk} = \Gamma_{ikj}$ (and hence $\Sigma_{ijk} = \Sigma_{ikj}$) one finds
\begin{align}
R'_{\alpha,i} &= R_{\alpha,i} + \varepsilon_{ik}(\mathbf{R}_{m}+\mathbf{R}_{\alpha})R_{\alpha,k}+ \omega_{ik}(\mathbf{R}_{m}+\mathbf{R}_{\alpha})R_{\alpha,k}\nonumber\\ 
& + [\varepsilon_{ik}(\mathbf{R}_{m}+\mathbf{R}_{\alpha}) - \varepsilon_{ik}(\mathbf{R}_{m})]R_{m,k}\nonumber\\ 
&\quad + [\omega_{ik}(\mathbf{R}_{m}+\mathbf{R}_{\alpha}) - \omega_{ik}(\mathbf{R}_{m})]R_{m,k}\nonumber\\ 
&\qquad-\left[\Sigma_{ijk}(\mathbf{R}_{m}+\mathbf{R}_{\alpha})-\Sigma_{ijk}(\mathbf{R}_{m})\right]R_{m,j}R_{m,k}\nonumber\\ 
&\qquad\quad - 2\Sigma_{ijk}(\mathbf{R}_{m}+\mathbf{R}_{\alpha})R_{\alpha,j}R_{m,k}\nonumber\\ 
&\qquad\qquad-\Sigma_{ijk}(\mathbf{R}_{m}+\mathbf{R}_{\alpha})R_{\alpha,j}R_{\alpha,k}.
\label{bond}
\end{align}
If the various strain contributions do not varies significantly on the scale of the bond length (which is require for Bloch theorem to hold locally) then $f(\mathbf{R}_{m}+\mathbf{R}_{\alpha}) \approx f(\mathbf{R}_{m}) $. Furthermore we can drop the last term in Eq. \eqref{bond} as it is second order the bond length and hence its contribution to the band structure will be small. Thus one finds
\begin{equation}
R'_{\alpha,i} = R_{\alpha,i} + \Omega_{ik}(\mathbf{R}_{m})R_{\alpha,k},
\end{equation}
with
\begin{equation}
\Omega_{ik}(\mathbf{R}_{m}) = \left\{\varepsilon_{ik}(\mathbf{R}_{m}) + \omega_{ik}(\mathbf{R}_{m}) - 2\Sigma_{ijk}(\mathbf{R}_{m})R_{m,j}\right\}.
\label{trans}
\end{equation}
The transformation $\Omega_{ik}(\mathbf{R}_{m})$ gives the displacement of the lattice vectors in the neighbourhood of $\mathbf{R}_{m}$ and is a function of the global coordinate, $\mathbf{R}_{m}$, only.

\section{Band Structure}

\begin{figure}[t]
\centering
\includegraphics[width=0.7\linewidth]{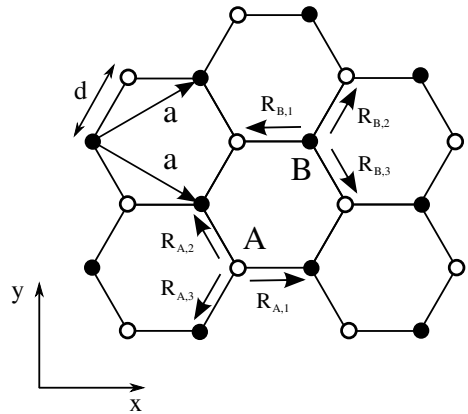}
\caption{The graphene lattice.}
\label{graphene}
\end{figure}

So far the discussion of strain has been general and can be applied to any material. Now we will consider graphene as an example. Graphene consists of two independent triangular sublattices (labelled $A$ and $B$). The unit cell is rhombic and contains two atoms, one from each sublattice, with nearest neighbour hopping connecting the two sublattices (See Fig. \ref{graphene}). 
The length of the lattice vector is $a = 2.46\mathrm{\AA}$ and the nearest neighbour vectors read
\begin{subequations}
\begin{align}
\mathbf{R}_{A,1} &= \left(\begin{array}{c}
\frac{a}{\sqrt{3}}\\
0
\end{array}\right), \hspace{0.2cm} 
\mathbf{R}_{A,2} = \left(\begin{array}{c}
-\frac{a}{2\sqrt{3}}\\
\frac{a}{2}
\end{array}\right), \hspace{0.2cm} 
\mathbf{R}_{A,3} = \left(\begin{array}{c}
-\frac{a}{2\sqrt{3}}\\
-\frac{a}{2}
\end{array}\right),\\
\mathbf{R}_{B,1}& = \left(\begin{array}{c}
-\frac{a}{\sqrt{3}}\\
0
\end{array}\right), \hspace{0.2cm} 
\mathbf{R}_{B,2} = \left(\begin{array}{c}
\frac{a}{2\sqrt{3}}\\
\frac{a}{2}
\end{array}\right), \hspace{0.2cm} 
\mathbf{R}_{B,3} = \left(\begin{array}{c}
\frac{a}{2\sqrt{3}}\\
-\frac{a}{2}
\end{array}\right),
\end{align}
\end{subequations}
each with length $d=a/\sqrt{3} = 1.42\,\mathrm{\AA}$.

The nearest-neighbour tight-binding Hamiltonian for each sublattice can be written as
\begin{equation}
\hat{H} = \left\{\sum_{\alpha}t_{\alpha}e^{-i\mathbf{k}\cdot[\mathbf{R}_{\alpha} + \bm{\Omega}(\mathbf{R}_{m})\cdot\mathbf{R}_{\alpha}]}\right\}\hat{a}^{\dagger}_{\mathbf{k}}\hat{b}_{\mathbf{k}} + \mathrm{h.c.},
\label{H}
\end{equation}
where the operators $\hat{a}^{(\dagger)}_{\mathbf{k}}$ and $\hat{b}^{(\dagger)}_{\mathbf{k}}$ create or annihilate electrons of momentum $\mathbf{k}$ from the $A$ and $B$ sublattices respectively. The sum over $\alpha$ is the sum over all nearest-neighbour vectors and $t_{\alpha}$ is the renormalized hopping amplitude. Much discussion has gone into the form of the hopping amplitude under strain. Here we use the parametrization \cite{pereiraTBM}, $t_{\alpha} = t_{0}\,\mathrm{exp}\left[-\beta\left(l/d-1\right)\right]$ where $l = |\mathbf{R}_{\alpha} + \bm{\Omega}(\mathbf{R}_{m})\cdot\mathbf{R}_{\alpha}|$ is the nearest neighbour distances under strain, $t_{0}\approx 2.8\,\mathrm{eV}$\cite{rev1} the unstrained hopping amplitude and $\beta \approx 3$ the hopping decay parameter \cite{pereiraTBM}. Since the displacement of the atoms is small we can expand Hamiltonian in Eq. \eqref{H} to linear order in strain. One finds that the tight-binding Hamiltonian becomes $\hat{H} = \hat{H}_{0} + \hat{H}_{\varepsilon}$ where $\hat{H}_{0}$ is the Hamiltonian for the unstrained graphene sheet and 
\begin{equation}
\hat{H}_{\varepsilon} = t_{0}\sum_{\alpha}e^{-i\mathbf{k}\cdot\mathbf{R}_{\alpha}}\left[\beta-\frac{\beta l}{d} - \mathbf{k}\cdot\bm{\Omega}(\mathbf{R}_{m})\cdot\mathbf{R}_{\alpha}\right]\hat{a}^{\dagger}_{\mathbf{k}}\hat{b}_{\mathbf{k}} + \mathrm{h.c.},
\label{Heps}
\end{equation}
is a strain-induced perturbation.

The usual expansion of the unstrained and the strain-induced perturbation Hamiltonians about the Dirac points at [$\mathbf{K}_{1,\pm}=(0,\pm 4\pi/3a)$, $\mathbf{K}_{2,\pm}=(\pm 2\pi/\sqrt{3}a,\pm 2\pi/3a), \mathbf{K}_{3,\pm}=(\pm 2\pi/\sqrt{3}a,\mp 2\pi/3a)$] leads to the usual linear band structure but with the replacement $\mathbf{k} \rightarrow \mathbf{k} - \mathbf{A}$, with the artificial vector potential, $\mathbf{A}$ at each Dirac point reading
\begin{subequations}
\begin{align}
\mathbf{A}_{1,\pm}(\mathbf{R}_{m}) &= \pm\frac{4\pi}{3a}\left(\begin{array}{c}
\Omega_{yx}\\
\Omega_{yy}
\end{array}\right)
\pm\mathbf{A}_{\beta}(\mathbf{R}_{m}),\label{A}\\\nonumber\\
\mathbf{A}_{2,\pm}(\mathbf{R}_{m}) &=\pm\frac{2\pi}{3a}\left(\begin{array}{c}
-\sqrt{3}\Omega_{xx}-\Omega_{yx}\\
-\sqrt{3}\Omega_{xy}-\Omega_{yy}
\end{array}\right)\pm\mathbf{A}_{\beta}(\mathbf{R}_{m}),\\
\mathbf{A}_{3,\pm}(\mathbf{R}_{m}) &=\pm\frac{2\pi}{3a}\left(\begin{array}{c}
\sqrt{3}\Omega_{xx}-\Omega_{yx}\\
\sqrt{3}\Omega_{xy}-\Omega_{yy}
\end{array}\right)\pm\mathbf{A}_{\beta}(\mathbf{R}_{m}),
\end{align}
\end{subequations}
where
\begin{equation}
\mathbf{A}_{\beta}(\mathbf{R}_{m}) = \frac{\sqrt{3}\beta}{2a}\left(\begin{array}{c}
\Omega_{xy}+\Omega_{yx}\\
\Omega_{xx}-\Omega_{yy}
\end{array}\right).\\
\end{equation}
The first term in the potential originates from the distortion of the lattice and the second from the renormalization of the hopping amplitude. Note these are slightly different from the expressions found in previous work \cite{pereiraA, geimhall,peeters1,peeters2} since in those studies the displacement is described solely by the strain tensor which is symmetric or by the strain and rotation tensors, the latter of which is antisymmetric. Thus, some simplification occurs. Here we have included the effect of the finite displacement term, which is asymmetric, and, therefore, have arrived at a more general expression. As the deformation, $\mathbf{\Omega}$, and, hence, the expressions for the artificial vector potentials are spatially varying they admit a curl and, therefore, describe a pseudo-magnetic field which near the $i$th Dirac point is given by $\mathbf{B}_{i,\pm} = \bm{\nabla}\times\mathbf{A}_{i,\pm}$.

\section{Point stretch of a graphene ribbon}

Here, we consider the point stretch of a wide graphene ribbon orientated such that the armchair edge is parallel to the $x$-axis and the zigzag edge is parallel to the $y$-axis. The ribbon is considered to be wide enough such that confinement effects are negligible and, hence, the band structure can be treated as that of bulk graphene. The ribbon is subject to a pair of equal and opposite point forces that act at opposing locations on the ribbon's edge. We define a coordinate system such that the origin is located at the centre of the ribbon, on the neutral, axis and the forces act along the $y$-axis at $x=0$ [See Fig. \ref{bandsplot} (a)]. 

One can show (see Appendix \ref{App:strain}) that the strain tensor for such a geometry is given by
\begin{equation}
\bm{\varepsilon}(x,y)  = \frac{F_{0}}{EL_{z}}\left(\begin{array}{cc}
-\frac{\nu}{(|x|+1)} & -\frac{\mathrm{sgn}\left[x\right]|y|}{2(|x|+1)^{2}}\\
-\frac{\mathrm{sgn}\left[x\right]|y|}{2(|x|+1)^{2}} & \frac{1}{(|x|+1)}
\end{array}\right),
\label{strain}
\end{equation}
where $F_{0}$ is the applied force, $L_{z}=3.5\,\mathrm{\AA}$ the thickness (in the $z$-direction) of the graphene ribbon \cite{graphmech} and $E\approx 340\,\mathrm{Nm}^{-1}$ ($\approx 1\,TPa$ acting over the graphene ribbon thickness, $L_{z}$) and $\nu=0.165$ are the Young's modulus \cite{graphmech} and poisson ratio \cite{poisson} of graphene, respectively. The coordinates $x\rightarrow x/L_{z}$ and $y\rightarrow x/L_{z}$ are dimensionless distances scaled by the thickness of the graphene ribbon. Note that the strain tensor for other orientation can easily be found via a rotational transformation. It is easy to show that the strain tensor satisfies the compatibility equation
\begin{equation}
\frac{\partial^{2} \varepsilon_{xx}}{\partial y^{2}} + \frac{\partial^{2} \varepsilon_{yy}}{\partial x^{2}}  - 2\frac{\partial^{2} \varepsilon_{xy}}{\partial x\partial y} = 0,
\end{equation}
and, hence, describes a unique, smoothly varying deformation. Furthermore, from $\bm{\nabla}\times\bm{\omega} = -\bm{\nabla}\times\bm{\varepsilon}$, one can show that the rotation tensor reads
\begin{equation}
\bm{\omega}(x,y)  = \frac{F_{0}}{EL_{z}}\left(\begin{array}{cc}
0& \frac{\mathrm{sgn}\left[x\right]|y|}{2(|x|+1)^{2}}\\
-\frac{\mathrm{sgn}\left[x\right]|y|}{2(|x|+1)^{2}} & 0
\end{array}\right).
\end{equation}
Using the form of the metric tensor given in Eq. \eqref{gij} one finds that the Christoffel Symbols are given by
\begin{subequations}
\begin{align}
\Gamma_{xxx} &= \frac{F_{0}}{EL_{z}}\frac{\nu}{(1+|x|)^2},\\
\Gamma_{yxx} &= \frac{F_{0}}{EL_{z}}\frac{2\,\mathrm{sgn}\left[x\right]|y|}{(1+|x|)^3},\\
\Gamma_{yxy} &= \Gamma_{yyx} = -\frac{F_{0}}{EL_{z}}\frac{1}{(1+|x|)^2},\\
\Gamma_{xyy} &= \Gamma_{xxy} = \Gamma_{xyx} = \Gamma_{yyy} = 0,
\end{align}
\end{subequations}
from which the components of $\bm{\Sigma}\cdot\mathbf{R}$ are calculated to be
\begin{multline}
\bm{\Sigma}\cdot\mathbf{R}(x,y) =\\
\frac{F_{0}}{EL_{z}}\left(\begin{array}{cc}
\nu\left[\frac{\mathrm{ln}\left[1+|x|\right]}{|x|}-\frac{1}{1+|x|}\right] & 0\\
\frac{|y|}{x}\left[\frac{\mathrm{ln}\left[1+|x|\right]}{|x|}-\frac{2|x|+1}{(1+|x|)^2}\right]& -\left[\frac{\mathrm{ln}\left[1+|x|\right]}{|x|}-\frac{1}{1+|x|}\right]
\end{array}\right).
\label{Sigma}
\end{multline}

\begin{figure*}[ht]
\centering
\includegraphics[width=1\linewidth]{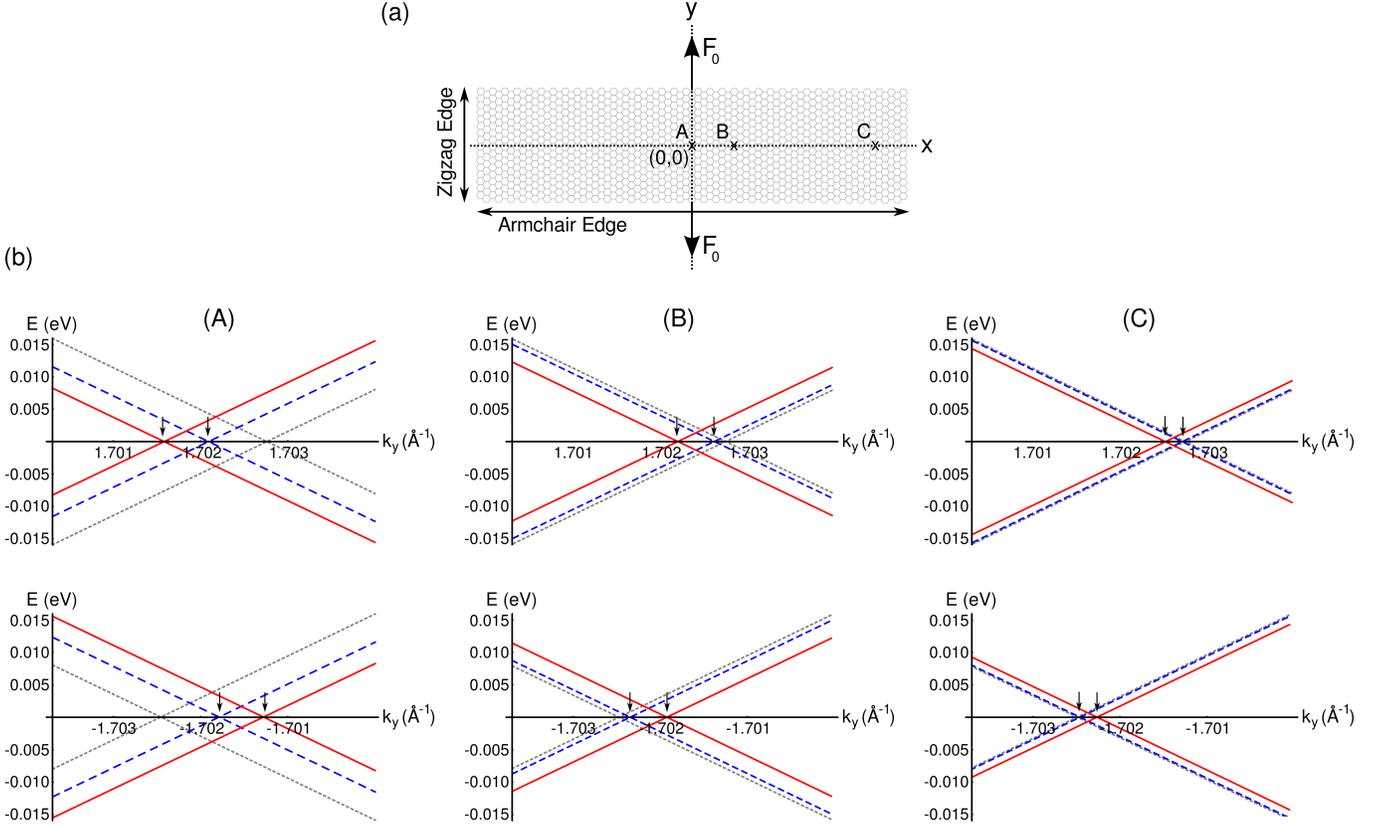}
\caption{(Color online) (a) Schematic diagram of the applied force. A pair of equal and opposing point forces,  $F_{0}$, act in the $\pm y$-directions at $x = 0$. (b) The local shift of the Dirac point for the positive (top) and negative (bottom) valleys for $y=0$ at (A) the origin, (B) $x = 10\,\mathrm{\AA}$ and (C) $x = 40\,\mathrm{\AA}$, respectively, for an applied force of $F_{0} = 0.35\,nN$. The red solid line shows the Dirac cone for the strained graphene lattice when the linear strain, rotation and finite displacement contributions are taken into account. The blue dashed line shows the Dirac cone for the strained graphene lattice when only the linear strain term is taken into account \cite{pereiraA}. The grey dotted line shows the Dirac cone for the unstrained graphene lattice. The arrows mark the artificial vector potential, $\mathbf{A}_{\mathbf{K}_{1,\pm}}$ as calculated by Eq. \eqref{A} (for the red curve) or the expressions found in previous studies \cite{pereiraA} (for the blue curve).}
\label{bandsplot}
\end{figure*}
Considering the band structure close to the $\mathbf{K}_{1,\pm}=(0,\pm 4\pi/3a)$ Dirac points and using the definition of the transformation $\mathbf{\Omega}$ from Eq. \eqref{trans}, one finds that the artificial vector potential that results from the strain is given by
\begin{subequations}
\begin{align}
&A_{1,\pm,x}=\nonumber\\
&\quad\mp\frac{F_{0}}{EL_{z}}\frac{8\pi + 3\sqrt{3}\beta}{6a}\frac{|y|}{x}\left[\frac{2\mathrm{ln}\left[1+|x|\right]}{|x|}-\frac{(2+3|x|)}{(1+|x|)^2}\right],\\
&A_{1,\pm,y}=\nonumber\\
&\quad\pm\frac{F_{0}}{EL_{z}}\frac{8\pi - 3\sqrt{3}\beta(1+\nu)}{6a}\left[\frac{2\mathrm{ln}\left[1+|x|\right]}{|x|}-\frac{1}{(1+|x|)}\right],
\end{align}
\end{subequations}
which leads to a pseudo-magnetic field of
\begin{multline}
B_{\mathbf{K}_{1,\pm},z}=\mp\frac{\hbar}{e}\frac{F_{0}}{EL_{z}^2}\frac{\sqrt{3}\beta}{2a}\frac{(2+\nu)}{x}\\
\times\left[\frac{2\mathrm{ln}\left[1+|x|\right]}{|x|}-\frac{(2+3|x|)}{(1+|x|)^2}\right].
\label{B}
\end{multline}

\begin{figure}[t]
\centering
\includegraphics[width=1\linewidth]{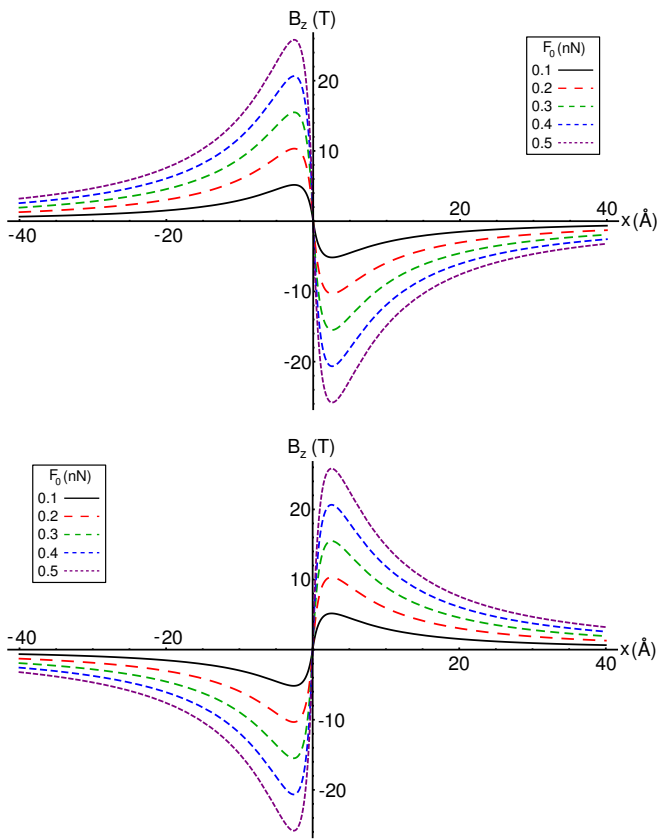}
\caption{(Color online) The pseudo-magnetic field for the positive (top) and negative (bottom) valleys for different applied forces.}
\label{bplot}
\end{figure}
A similar calculation shows that the expressions for the pseudo-magnetic fields at $\mathbf{K}_{2,\pm}$ and $\mathbf{K}_{3,\pm}$ are identical. Note that the field is constant in the $y$-direction and hence does not change over the height of the ribbon. Considering the form of the pseudo-magnetic field given in Eq. \eqref{B} one can see that it is proportional to $\beta$ and hence is a result of the renormalization of the hopping amplitude only. In this particular case, the distortion of the lattice does not contribute. This is consistent with previous studies \cite{error, peeters2}. Furthermore, comparing, Eq. \eqref{B} with Eq. \eqref{Sigma} one can see that the pseudo-magnetic field is solely the result of the finite displacement term, $\bm{\Sigma}\cdot\mathbf{R}$.

Figure \ref{bandsplot} (b) shows the local band structure near the Dirac point for the $\mathbf{K}_{1,\pm}$ valleys at various points along the graphene ribbon. One can see that the Dirac cone is shifted compared to unstrained graphene, $\mathbf{k}\rightarrow\mathbf{k}-\mathbf{A}$, and hence an artificial vector potential has been induced by the strain. The shift (artificial vector potential) is largest close to the origin, where the point forces act, and falls off as one moves away from the origin. At large distances the location of the Dirac cone approaches that of unstrained graphene. This spatial variation of the artificial vector potential gives rise to the pseudo-magnetic field. Note that, by taking the finite displacement terms into account one sees that, at the origin, the artificial vector potential is almost double that which is predicted by the linear strain theory alone.

Figure \ref{bplot} shows the pseudo-magnetic field for various applied forces. One sees a `domain wall' structure with a sudden change in the orientation of the field at the origin. Increasing the applied force increases the maximum field, increasing the height of the domain wall. By taking the rotation and finite displacement terms into account one sees that the strength of the field is over three times that which is predicted by the linear strain theory (a $0.1\,nN$ gives a field of $\approx 5\,T$ compared to the $\approx 1.5\,T$ field given by linear strain theory \cite{pereiraA}). Similarly, we see up to a factor of two increase in the pseudo-magnetic field strength compared to studies which used finite strain theory and included the rotation tensor (strains of $\approx 12\,\%$ lead to fields of $\approx 1500\,T$ where as previous studies find fields of $800-1400\,T$ depending on the strain profile \cite{peeters2}). Finally, pseudo-magnetic field profiles have been predicted for strain induced via nanostructured substrates \cite{peeters1,peeters3,peeters4}, however, this studies shows it is possible to create similar field profiles with in-plane strains only.

In addition to the appearance of a pseudo-magnetic field, some studies have predicted the appearance of pseudo-scalar fields which are linearly proportional to the average change in the bond length \cite{scalar1,scalar2,scalar3}. The finite displacement term contributes to the change in bond length and, hence, changes the pseudo-scalar potential. The change in bond length depends greatly of the strength and direction of the applied forces and the location of the bond in the material, however, in the current study the change in bond length can be as large as a factor of two. 

Finally, it is worth comparing the above analytical results with those obtained via atomistic simulations. Such first principle simulations should not suffer from the approximations imposed by linear strain theory. The current study has considered low strains of $\approx 0.25\,\%$ comparable to previous studies of linear strain theory \cite{pereiraA}. However, the theory is valid for large strains as well with strains of $\approx 12\,\%$ lead to pseudo-magnetic fields of $\approx 1500\,T$. Out of plane deformations with similar magnitude strain fields have been studied previously using atomistic simulations \cite{peeters1,peeters3} and the calculated pseudo-magnetic field is of similar magnitude to those predicted by the current study.

\section{Summary}

We have developed a theory of strain induced band structure engineering that goes beyond the small displacement, infinitesimal vector transformations of linear and finite strain theory. By integrating the infinitesimal strain tensor over the deformation, and thereby finding the strain transformation of finite vectors under large displacements, we obtain a finite displacement term which gives a significant contribution to the strained band structure. Further to this, we found that a point stretch of a wide graphene ribbon by a force on the order of $0.1\,nN$ generates a `domain wall' like magnetic field profile with field strength on the order of $5\,T$, over three times as much as predicted by the linear strain theory. This ability to generate and tailor complex pseudo-magnetic field structures allows for unprecedented control of the electrons in graphene and could pave the way for many novel magneto-electronic and spintronic devices.

\section{Acknowledgements}

The author would like to thank P. Del Linz for useful discussions and A. Danner and the members of the Optical Device Research Group at the National University of Singapore for their hospitality. The author would also like to thank F. M. Peeters for bring the latest developments in the field to his attention.

\appendix

\section{Strain tensor of a point stretch} 
\label{App:strain}

\begin{figure}[htb]
\centering
\includegraphics[width=1\linewidth]{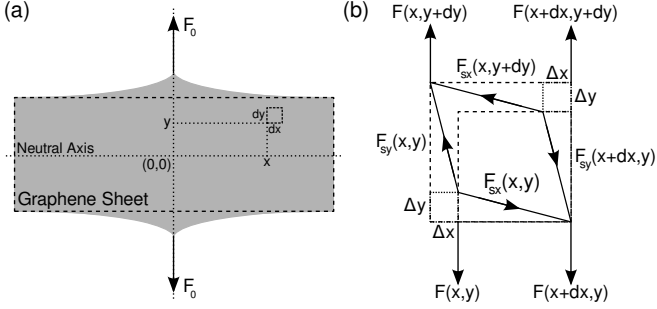}
\caption{(a) Schematic diagram of the applied force. (b) Force balance of an area element $(dx,dy)$ at point $(x,y)$.}
\label{stretch}
\end{figure}

Here we derive the strain tensor for a material ribbon under a point stretch. The ribbon is considered to be unbounded in the $x$ direction but bounded in the $y$-direction. The ribbon is subject to a pair of equal and opposite point forces that act at opposing locations on the ribbon's edge. We define a coordinate system such that the origin is located at the centre of the ribbon, on the neutral axis, and the forces act along the $y$-axis at $x=0$. The stress tensor can be computed by performing a force balance on an infinitesimal area element at a location $(x,y)$ relative to the origin (See Fig. \ref{stretch}).
For a static element both the tensile stresses and moments must vanish. In the following we will consider the upper right quadrant ($+x$, $+y$) of the coordinate system. The other three quadrants follow identically with an appropriate change of sign.

The $\sigma_{xx}$ component vanishes trivially as there is no component of the force acting in the $x$-direction. The $\sigma_{yy}$ component can be found from a moment balance (the tensile stress balance in the $y$-direction gives a trivial constraint)
\begin{equation}
F(x,y)x = F(x+dx,y)[x+dx] + L_{z}[F_{sx}(x,y+dy) - F_{sx}(x,y)],
\end{equation}
which to first order gives
\begin{equation}
x\frac{dF(x,y)}{dx}dx+F(x,y)dx + L_{z}\frac{dF_{sx}(x,y)}{dy}dy=0,
\label{moment}
\end{equation}
where $F_{sx}(x,y)$ is the shear stress along the length, $dx$, and across thickness ($z$-direction), $L_{z}$, of the ribbon. This term appears because shear stress is able to transfer moment laterally. The shear stress in the $x$ direction must equal the difference in tensile stress at $x$ and $x+dx$
\begin{equation}
\frac{dF_{sx}(x,y)}{dy}dy = F(x+dx,y) - F(x,y) = \frac{dF(x,y)}{dx}dx.
\label{shear}
\end{equation}
Substituting the definition in Eq. \eqref{shear} into Eq. \eqref{moment} leads to
\begin{equation}
(x+L_{z})\frac{dF(x,y)}{dx}+F(x,y) = 0,
\end{equation}
which has the solution
\begin{equation}
F(x,y) = \frac{c}{(x+L_{z})}.
\end{equation}
At $x=0$ the stress must be equal to the applied force acting over the thickness. Thus, $c=F_{0}$ where $F_{0}$ is the applied force at the ribbon edge.

The shear stress can be found from Eq. \eqref{shear},
\begin{align}
\frac{dF_{sx}(x,y)}{dy}dy &= \frac{dF(x,y)}{dx}dx,\nonumber\\
\frac{dF_{sx}(x,y)}{dy} &= \frac{dF(x,y)}{dx}\frac{dx}{dy},
\end{align}
where, owing to shear deformation, we have
\begin{align}
\frac{dx}{dy} &= \frac{\Delta x}{\Delta y} = G,
\end{align}
where $G$ is the shear modulus, which, for a planar orthotropic material, is given in terms of the poisson ratio, $\nu$, by $G=1/2(1+\nu)$. Thus
\begin{align}
\frac{dF_{sx}(x,y)}{dy} &= \frac{1}{2(1+\nu)}\frac{dF(x,y)}{dx},\nonumber\\
\frac{dF_{sx}(x,y)}{dy} &= -\frac{1}{2(1+\nu)}\frac{F_{0}}{(x+L_{z})^{2}},\nonumber\\
F_{sx}(x,y) &= -\frac{1}{2(1+\nu)}\frac{F_{0}y}{(x+L_{z})^{2}}+c.
\end{align} 
On the neutral axis ($y=0$) the shear stress vanishes, hence $c=0$. Finally, the stress tensor over all four quadrants is given by
\begin{equation}
\bm{\sigma}(x,y) = \frac{F_{0}}{L_{z}}\left(\begin{array}{cc}
0 & -\frac{1}{2(1+\nu)}\frac{\mathrm{sgn}\left[x\right]|y|}{(|x|+1)^{2}}\\
-\frac{1}{2(1+\nu)}\frac{\mathrm{sgn}\left[x\right]|y|}{(|x|+1)^{2}} & \frac{1}{(|x|+1)}
\end{array}\right),
\end{equation}
where $x\rightarrow x/L_{z}$ and $y\rightarrow x/L_{z}$ have been scaled by the ribbon thickness.

The components of the strain tensor can be found from the compliance tensor. For a planar, isotropic material, the transformation reads
\begin{equation}
\left(\begin{array}{c}
\varepsilon_{xx}(x,y)\\
\varepsilon_{yy}(x,y)\\
\varepsilon_{xy}(x,y)\\
\end{array}\right) = 
\frac{1}{E}\left(\begin{array}{ccc}
1 & -\nu & 0 \\
-\nu & 1 & 0\\
0 & 0 & (1+\nu)
\end{array}\right)
\left(\begin{array}{c}
\sigma_{xx}(x,y)\\
\sigma_{yy}(x,y)\\
\sigma_{xy}(x,y)\\
\end{array}\right),
\end{equation}
where, $E$ is the Young's modulus which here carries units of $Nm^{-1}$. Thus, the strain tensor reads
\begin{equation}
\bm{\varepsilon}(x,y)  = \frac{F_{0}}{EL_{z}}\left(\begin{array}{cc}
-\frac{\nu}{(|x|+1)} & -\frac{\mathrm{sgn}\left[x\right]|y|}{2(|x|+1)^{2}}\\
-\frac{\mathrm{sgn}\left[x\right]|y|}{2(|x|+1)^{2}} & \frac{1}{(|x|+1)}
\end{array}\right).
\end{equation}

%%%%%%%%%%%%%%%%%%%%%%%%%%%%%%%%%%%%%%%%%%%%%%%%%%%%%%%%%%%%%%%%%%%%%%

\end{document}